\documentclass[structabstract]{aa}

\usepackage{txfonts}
\usepackage{graphicx}
\usepackage{epsfig}
\usepackage{color}
\usepackage{natbib}
\bibpunct{(}{)}{;}{a}{}{,} 

\begin{document}

\title{Variability and stellar populations with deep optical-IR
images of the Milky Way disc: matching VVV with VLT/VIMOS data\thanks{Based
on observations collected with the VLT and VISTA telescopes at Paranal
Observatory (ESO Programmes 075.C-0427(A) and 179.B-2002(B),
respectively)}$^,$\thanks{Photometry and data table on detected variables
are available in electronic form at the CDS via anonymous ftp to
cdsarc.u-strasbg.fr (130.79.128.5) or via
http://cdsweb.u-strasbg.fr/cgi-bin/qcat?J/A+A/000/000}
}
\author{P. Pietrukowicz\inst{1}
        \and
        D. Minniti\inst{2,3}
        \and
        J. Alonso-Garc\'ia\inst{2}
        \and
        M. Hempel\inst{2}
        }

\institute{Warsaw University Observatory, Al. Ujazdowskie 4,
           00-478 Warszawa, Poland\\
\email{pietruk@astrouw.edu.pl}
       \and
Departamento de Astronom\'ia y Astrof\'isica,
Pontificia Universidad Cat\'olica de Chile, Av. Vicu\~na MacKenna 4860,
Casilla 306, Santiago 22, Chile
       \and
Vatican Observatory, Vatican City State V-00120, Italy
}

\titlerunning{Variability and stellar populations in the Milky Way disc}
\subtitle{}
\authorrunning{Pietrukowicz et al.}

\date{Received -; accepted -}

\abstract
{}
{
We have used deep $V$-band and $JHK_s$-band observations to investigate
variability and stellar populations near the Galactic plane in Centaurus,
and compared the observations with the Galactic model of Besan\c{c}on.
}
{
By applying image subtraction technique to a series of over 580 $V$-band
frames taken with the ESO VLT/VIMOS instrument during two contiguous nights
in April 2005, we have detected 333 variables among 84~734 stars in the
brightness range $12.7<V<26.0$~mag. Infrared data collected in March 2010
with the new ESO VISTA telescope allowed us to construct deep combined
optical-IR colour-magnitude and colour-colour diagrams.
}
{
All detected variables but four transit candidates are reported for the
first time. The majority of the variables are eclipsing/ellipsoidal
binaries and $\delta$~Scuti-type pulsators. The occurrence rate of
eclipsing/ellipsoidal variables reached $\sim$0.28\% of all stars. This
is very close to the highest fraction of binary systems detected using
ground-based data so far (0.30\%), but still about four times less
than the average occurrence rate recently obtained from the $Kepler$ space
mission after 44~days of operation. Comparison of the observed $K_s$ vs.
$V-K_s$ diagram with a diagram based on the Besan\c{c}on model shows
significant effects of both distance and reddening in the investigated
direction of the sky. We demonstrate that the best model indicates
the presence of absorbing clouds at distances 11-13~kpc from the Sun
in the minor Carina-Sagittarius Arm.
}
{}


\maketitle

\section{Introduction}

In recent years our knowledge on the structure of the Milky Way
Galaxy has greatly improved thanks to wide-field photometric surveys
conducted from ground-based and orbital telescopes. They are usually
dedicated to the detection of particular objects, such as transiting
extrasolar planets, microlensing events, or gamma ray burst afterglows,
e.g., MACHO \citep{alco00}, OGLE \citep{uda03}, HAT \citep{bak04},
ROTSE \citep{woz04}, ASAS \citep{poj01}, and the {\it Kepler} space
mission \citep{koch10}. However, these surveys also discover
an enormous amount of new variable objects of different types
\citep[e.g.,][]{woz02,poj05,nat10}. Variable stars provide important
information about the structure and evolution of stars as well
as the whole Galaxy \citep[e.g.,][]{alco98,szcz09}.

The most recent view of the Galactic plane regions has been presented by
\cite{chu09} based on the {\it Spitzer}/GLIMPSE surveys.
The large-scale structure of the Galaxy has been traced by red-clump
giants to reveal the radius and orientation of the central bar,
thanks to characteristic changes in star counts for $|l|<65$\degr
as a function of Galactic longitude at a wavelength of $4.5$~$\mu$m.

A useful method to investigate structure and stellar populations
in the Galaxy is the comparison of synthetic magnitude histograms,
colour-magnitude diagrams (CMDs), and colour-colour diagrams with real data
\citep{luc08}. Here, we construct deep
(down to $V\approx26.0$~mag and $K_s\approx18.6$~mag) optical-IR
colour-colour and colour-magnitude diagrams of two southern Galactic
plane fields. For one of these fields we present the results of a search
for variable stars and compare the observed $K_s$ vs. $V-K_s$ diagram
with a theoretical one based on the Galactic model of Besan\c{c}on
\citep{rob03}. Our results show the great potential of ongoing
and upcoming large visual and near-infrared surveys, such as LSST
\citep[Large Synoptic Survey Telescope,][]{ive08}, VVV \citep[Vista
Variables in Via Lactea,][]{min10}, or Pan-STARRS \citep{kai02}.

\section{Observations and data reduction}

The observations presented in this work were carried out with the ESO VISTA
and UT3 telescope at Paranal Observatory. The infrared observations were
collected with the new 4.1-m Visible and Infrared Survey Telescope for Astronomy
(VISTA) as a part of the VISTA Variables in the Via Lactea (VVV) ESO Public
Survey \citep{min10}. The telescope is equipped with a $16\times2048\times2048$
pixel near-IR camera with a pixel size of $0\farcs34$. The monitored area
of the VVV survey covers the Galactic bulge ($-10\degr<l<+10.5\degr$,
$-10\degr<b<+5.6\degr$) and an adjacent section of the Galactic plane
($-65\degr<l<-10\degr$, $-2.25\degr<b<+2.25\degr$). Planned observations
span the years 2010-2014 with the main variability monitoring campaigns
in 2012 (bulge area) and 2013 (disc area). We used the DOPHOT package
\citep{sch93} to extract photometry from images obtained with VISTA.

Optical data were obtained with VIMOS at the 8.2-m Unit Telescope 3 (UT3)
of the Very Large Telescope in April 2005. VIMOS is an imager and multi-object
spectrograph \citep{lef03}. Its field of view consists of four quadrants of about
$7\arcmin \times 8\arcmin$ each, separated by a cross, $2\arcmin$ wide.
Each CCD has $2048 \times 2440$ pixels with a pixel size of $0\farcs205$.
The main goal of that programme was to perform a photometric follow-up of
over 30 OGLE transiting candidates spread over four VIMOS fields in the
constellations Carina, Centaurus and Musca. Some individual results
were published by \cite{fer06}, \cite{diaz07}, \cite{hoy07}, \cite{min07},
\cite{pie09} and \cite{pie10}.

The work presented here focuses on two VIMOS fields located within
the VVV area in Centaurus. They contain OGLE-TR-167 and OGLE-TR-170,
both classified as planetary transit candidates by \cite{uda04},
but currently known as eclipsing systems \citep{pie10}. We will
refer to the fields as F167 and F170, respectively. They are located
within the VVV tiles\footnote {For the definition of the tile
see the VVV description paper by \cite{min10}} d009 and d008,
respectively. The field F167 was monitored for two hours in the night
of 2005 April 9, and during the whole night of 2005 April 10, while the field
F170 was observed for the whole two subsequent nights of 2005 April 11-12.
All VIMOS images were taken in the $V$-band only. Three hundred ninty-five and 583
exposures were obtained in the fields F167 and F170, respectively.
Table~\ref{tab:fields} gives basic information on these fields.

\begin{table*}[ht]
\centering
\caption{Characteristics of the observed fields. Coordinates are given for the
centres of the fields. Reddening values were taken from \cite{sch98}.}
\smallskip
{\small
\begin{tabular}{ccccccc}
VIMOS field  & VVV field & RA(2000.0) & Dec(2000.0) & $l$ & $b$ & $E(B-V)$ \\
 & & [h:m:s] & [$\degr$:$\arcmin$:$\arcsec$] & [$\degr$] & [$\degr$] & [mag] \\
\hline
F167 & d009 & 13:31:36.00 & -64:04:15.0 & 307.306 & -1.541 & $1.7-2.1$ \\
F170 & d008 & 13:14:17.60 & -64:44:21.0 & 305.368 & -1.976 & $1.0-1.4$ \\
\hline
\label{tab:fields}
\end{tabular}}
\end{table*}

\begin{figure}
\includegraphics[angle=0,width=0.5\textwidth]{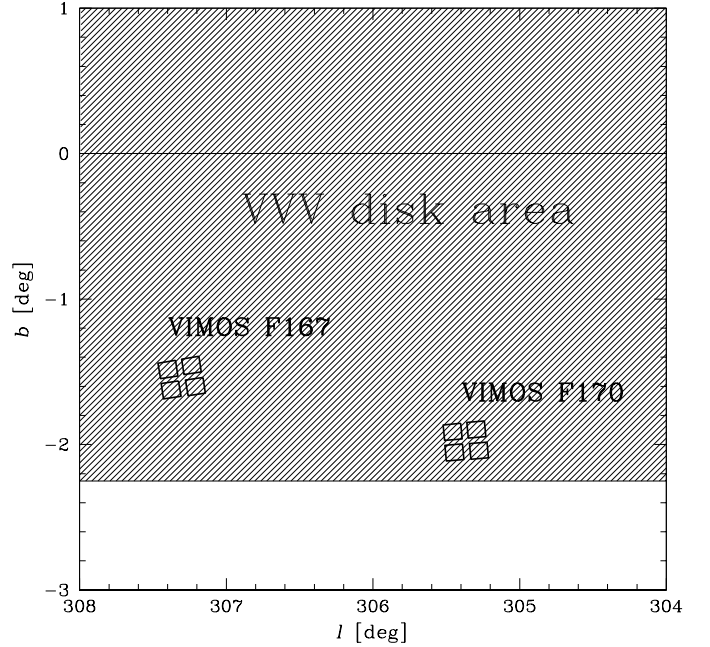}
\caption{Location of the analysed fields in Galactic coordinates.}
\label{fig:map}
\end{figure}

The periphery of each VIMOS quadrant suffers from coma. Therefore,
we reduced the field of view to $1900 \times 2100$ pixels, which then
covered $7\farcm18 \times 6\farcm49$. The total field in which we searched
for variable objects equals 186.3 arcmin$^2$. In Fig.~\ref{fig:map} we show
the location of the two VIMOS fields overlaid on a part of the VVV disc area.

Owing to the longer time span and larger number of images taken
in the field F170, we decided to focus our search in this field.
The photometry was extracted with the help of the
{\it Difference Image Analysis Package} (DIAPL) written by \cite{woz00}
and modified by W. Pych\footnote{The package is available at
http://users.camk.edu.pl/pych/DIAPL/}. The package is an implementation
of a method developed by \cite{ala98}. To obtain higher quality
photometry, we divided the field into $475 \times 525$ pixel subfields.

\begin{figure}[htb]
\includegraphics[angle=0,width=0.5\textwidth]{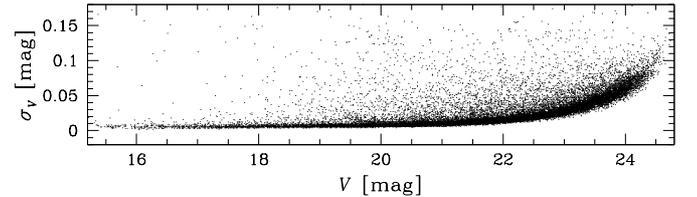}
\caption{Photometric errors for 22~717 stars detected in the VIMOS
quadrant A2 of the field F170 plotted as a function of $V$-band magnitude.}
\label{fig:sigma}
\end{figure}

Reference frames were constructed by combining nine of the highest
quality individual images, i.e. with the best seeing and low background.
The profile photometry for the reference frame was extracted
with DAOPHOT/ALLSTAR \citep{stet87}. These measurements
were used to transform the light curves from differential flux units into
instrumental magnitudes, which were subsequently transformed into standard
$V$-band magnitudes by adding an offset derived from $V$-band
magnitudes of the planetary transits located in the field.
The quality of the photometry is illustrated in Fig.~\ref{fig:sigma}.

\section{The variables}

Owing to the relatively short period of the VIMOS observations in the field F170
(only two contiguous nights), we decided to look for variables
only by direct eye inspection of all 84~734 extracted light curves.
The total number of detected variables reached 333 objects. All objects, except
OGLE-TR-170, OGLE-TR-171 \citep{uda04}, and transit-3 and transit-4 \citep{pie10}
are new identifications. We used no automatic methods for the search.
Previous experience with the four-night data in the VIMOS field F113 towards
the constellation Carina \citep{pie09} showed us that about one third
of the periodic variables were missed in the automatic search and merely four
additional objects were found in comparison to the simple eye inspection.
For the two-night data in Centaurus this would be even less effective.

\begin{table}[hb]
\caption{Census of variables detected in the VIMOS field F170.}
\smallskip
{\small
\begin{tabular}{lrr}
\hline
Type of stars                          & Number & Percentage \\
\hline
All stars searched for variability      & 84734 &   100 \% \\
\hline
All variables                           &   333 & 0.393 \% \\
\hline
All binaries                            &   241 & 0.284 \% \\
Eclipsing/ellipsoidal with known period &   210 & \\
Transiting                              &     4 & \\
Other eclipsing                         &    27 & \\
\hline
Pulsating variables                     &    53 & 0.063 \% \\
$\delta$ Scuti                          &    47 & 0.055 \% \\
Other pulsating                         &     6 & \\
\hline
Long-time scale variables               &    38 & 0.045 \% \\
\hline
Stars with flares                       &     2 & 0.002 \% \\
\hline
\label{tab:variables}
\end{tabular}}
\end{table}

All detected variables were sorted by increasing right ascension and
classified by considering the shape of the light curves and possible
periodicity. In Table~\ref{tab:variables} we provide a census of the different types of
variables found in field F170. It is worth to compare this table with Table~2
in \cite{pie09}, which was prepared for the field F113, located at similar
galactic latitude but $\sim$16$\degr$ farther away from the Galactic centre.
The lower percentage of detected variables, 0.393\% vs. 0.692\%, can be
explained by the shorter time coverage of the observations (two nights vs. four nights).
An interesting result is the very similar percentage of eclipsing/ellipsoidal
variables in both samples: 0.284\% in F170 vs. 0.291\% in F113. We were able
to confirm the periodic nature of 263 stars out of 329 newly discovered variables.
We derived the periods of 210 out of 237 new eclipsing/ellipsoidal
binary systems. The incident rate of observed
$\delta$~Scuti-type pulsators in field F170 is about three times
lower than in field F113. The detection of this type of low-amplitude variables
strongly depends on their distance. The majority of the stars in field
F170 are objects located in the Scutum-Centaurus Arm, while most
of the stars observed in field F113 are likely located closer to
us in the minor Carina-Sagittarius Arm.

\begin{figure*}[htb]
\includegraphics[angle=0,width=1.0\textwidth]{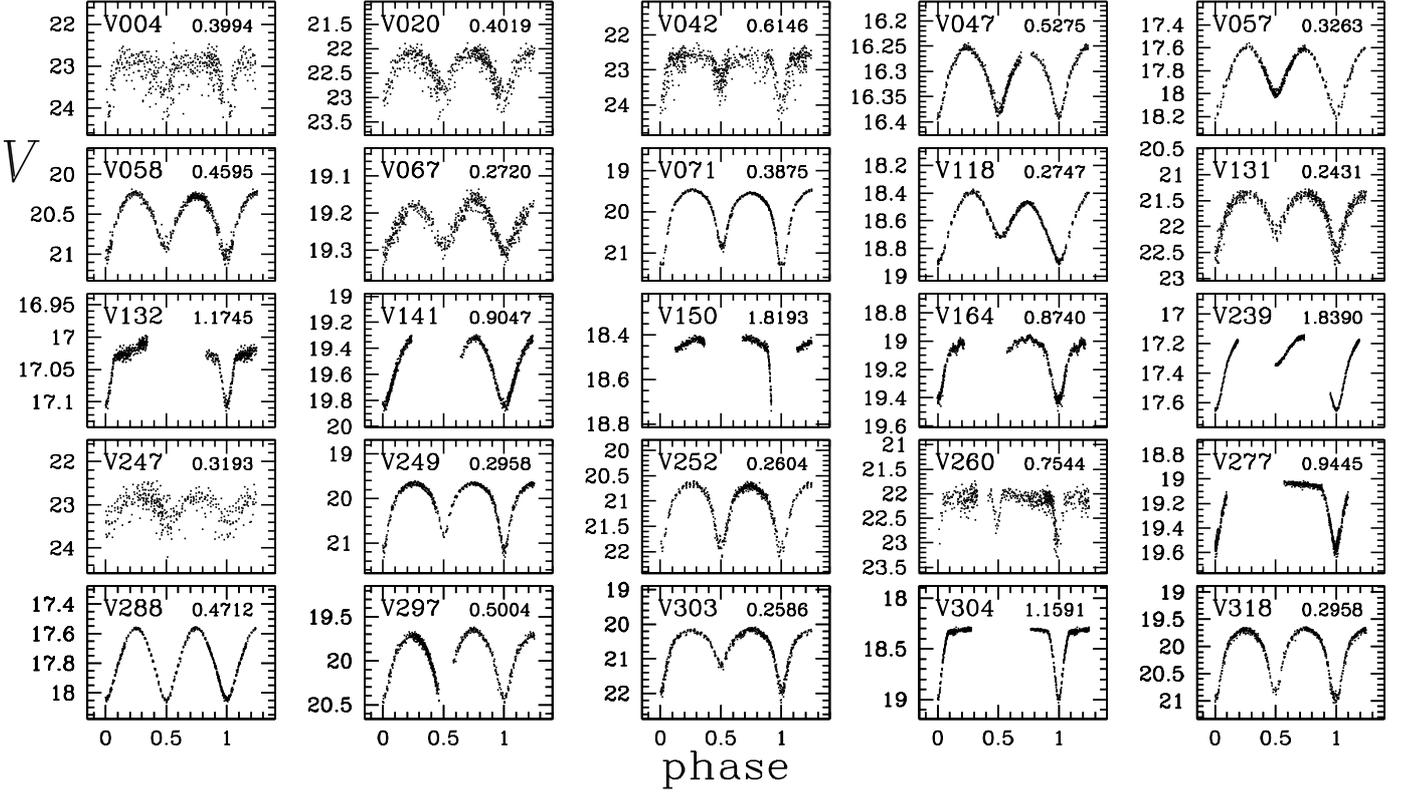}
\caption{Examples of phased light curves of eclipsing/ellipsoidal variables
with estimated period (in days). Note that in object V014 we observed
a flare of an amplitude of 0.1~mag.}
\label{fig:lceclper}
\end{figure*}

\begin{figure*}[htb]
\includegraphics[angle=0,width=1.0\textwidth]{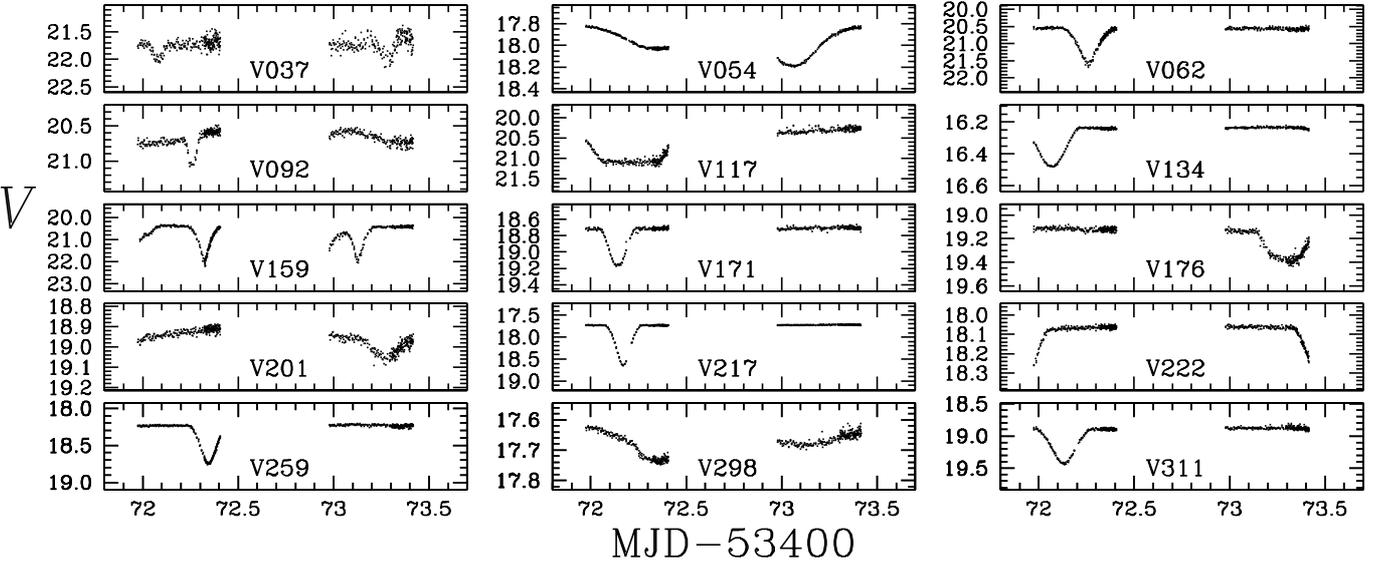}
\caption{Examples of light curves of eclipsing variables with unknown period.}
\label{fig:lcecllon}
\end{figure*}

\begin{figure*}[htb]
\includegraphics[angle=0,width=1.0\textwidth]{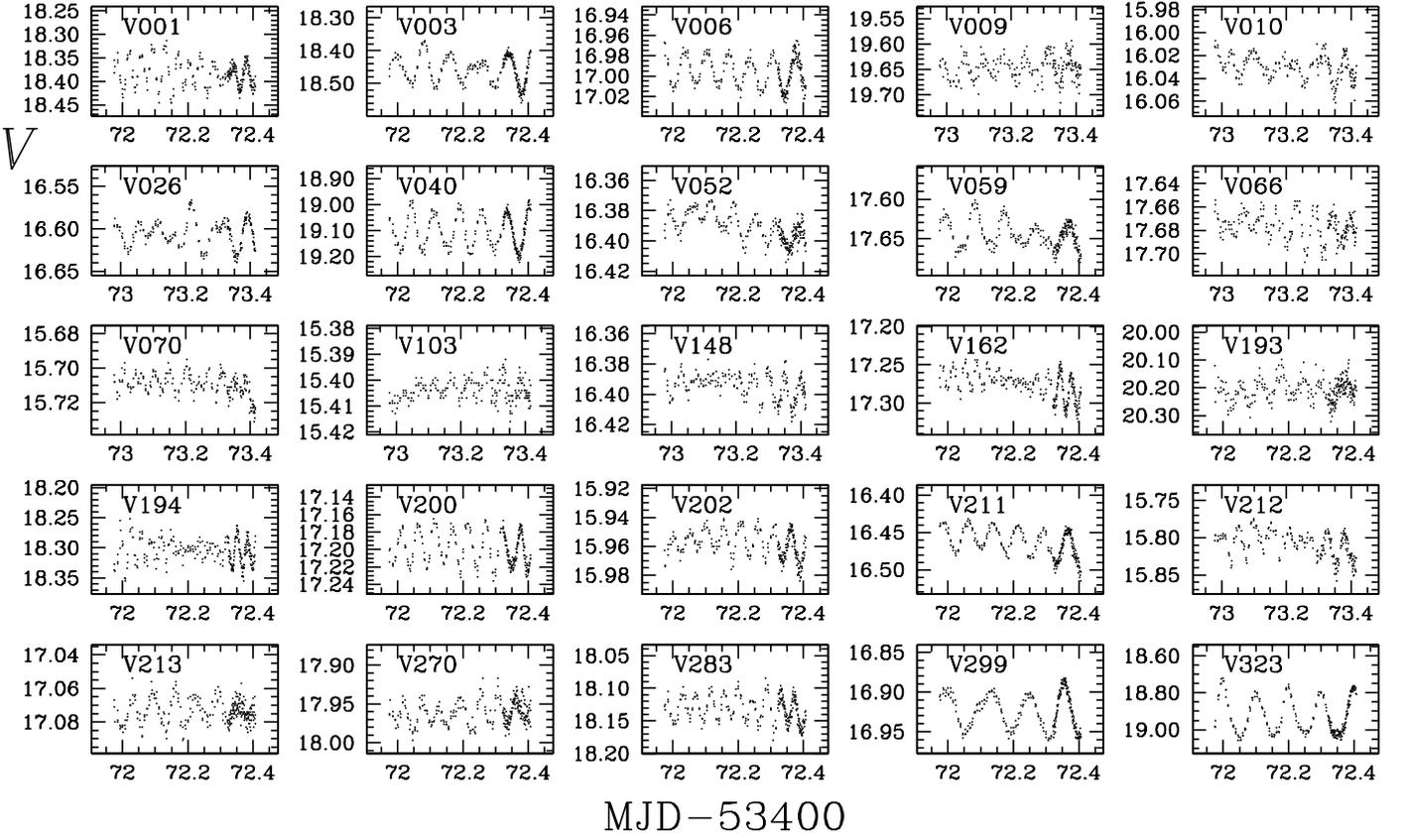}
\caption{Examples of light curves of detected $\delta$~Scuti-type variables.
Each panel presents data points from a single night only.}
\label{fig:lcdsc}
\end{figure*}

\begin{figure*}[htb]
\includegraphics[angle=0,width=1.0\textwidth]{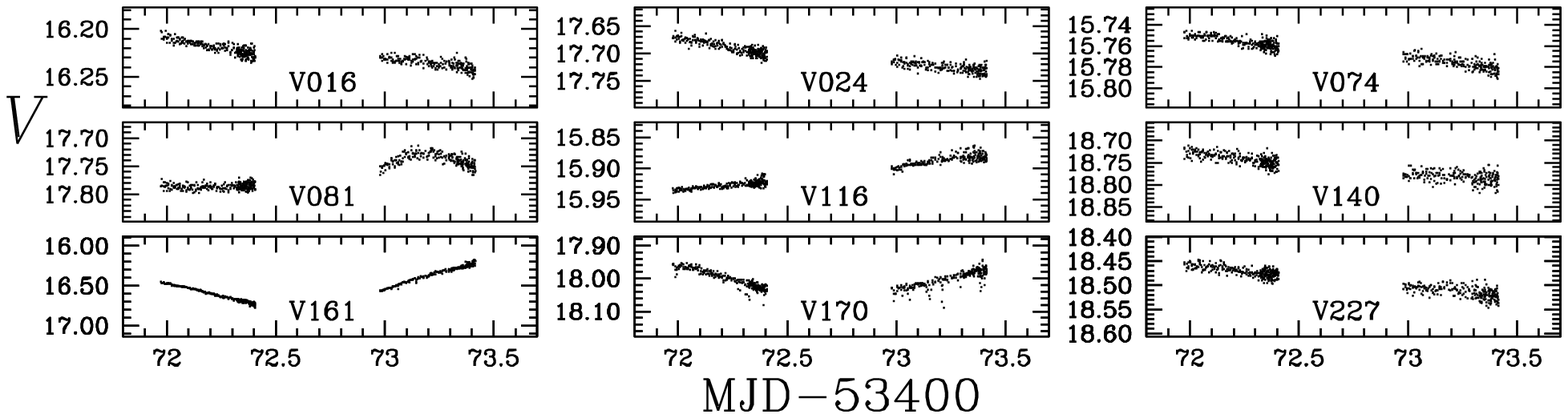}
\caption{Light curves of miscellaneous variables of long-time
scale brightness changes.}
\label{fig:lclon}
\end{figure*}

In Figs.~\ref{fig:lceclper} to~\ref{fig:lclon} we present some example light
curves of the variables found in the VIMOS field F170. The eclipsing binaries
V004, V042, and V247 in Fig.~\ref{fig:lceclper} are among the faintest
variables in the sample. The detection of a flare in the light curve
of star V014 is one of the more peculiar results. Similarly we note
the rising part in the light curve of the eclipsing object V159
in Fig.~\ref{fig:lcecllon}. Close inspection showed that it is an
isolated star located far from any edges and defects of the VIMOS chip.
Most of the long-time scale variables illustrated in Fig.~\ref{fig:lclon} exhibit
monotonic changes in brightness. Object V081 presents an interesting light curve
with a flat curve in the first night followed by a smooth up and down
0.04~mag amplitude variation on the second night. Additional photometry
would help to classify these objects and to estimate the periods more accurately.

All photometric data presented in this paper, finding charts,
and a large table with coordinates, periods, and magnitudes of the
detected variables are available via anonymous ftp to
ftp.astrouw.edu.pl/pub/pietruk/VIMOSvar/ or cdsarc.u-strasbg.fr.

\section{Matching optical and infrared data}

We have matched the $V$-band data obtained with VIMOS for the fields
F167 and F170 with $JHK_s$-band observations from VISTA using equatorial
coordinates. Prior to this operation we prepared auxiliary maps showing
positions of stars from both lists and we applied position corrections
in each VIMOS quadrant. Because of the large difference in brightness of stars
in the optical and infrared passbands and the effect of coma in the
corners of the VIMOS detectors we decided to set a matching radius
of 1\farcs2, and to reject all ambiguous objects from the final list.

\subsection{Star count histograms}

Table~\ref{tab:allstars} gives numbers of all detected and matched
stars together with magnitude range in the $V$ and $K_s$ bands.
The difference of 0.8~mag in the $K_s$-band limit for the brightest
stars between the two fields results from different saturation values
of chips of the infrared camera at the VISTA telescope.

\begin{table}[h!]
\caption{Census and magnitude range of all analysed stars in
fields F167 and F170.}
\smallskip
{\small
\begin{tabular}{lrr}
\hline
Field (VIMOS-VVV)                  & F167-d009 & F170-d008 \\
\hline
All stars detected in the $V$ band     & 89694 & 84734 \\
All stars detected in the $K_s$ band   & 43852 & 62460 \\
Common stars                           & 28710 & 42813 \\
\hline
$V$-band range [mag]               & 12.5-25.9 & 12.7-26.0 \\
$K_s$-band range [mag]             & 13.2-18.6 & 12.4-18.6 \\
\hline
\label{tab:allstars}
\end{tabular}}
\end{table}

\begin{figure}
\includegraphics[angle=0,width=0.5\textwidth]{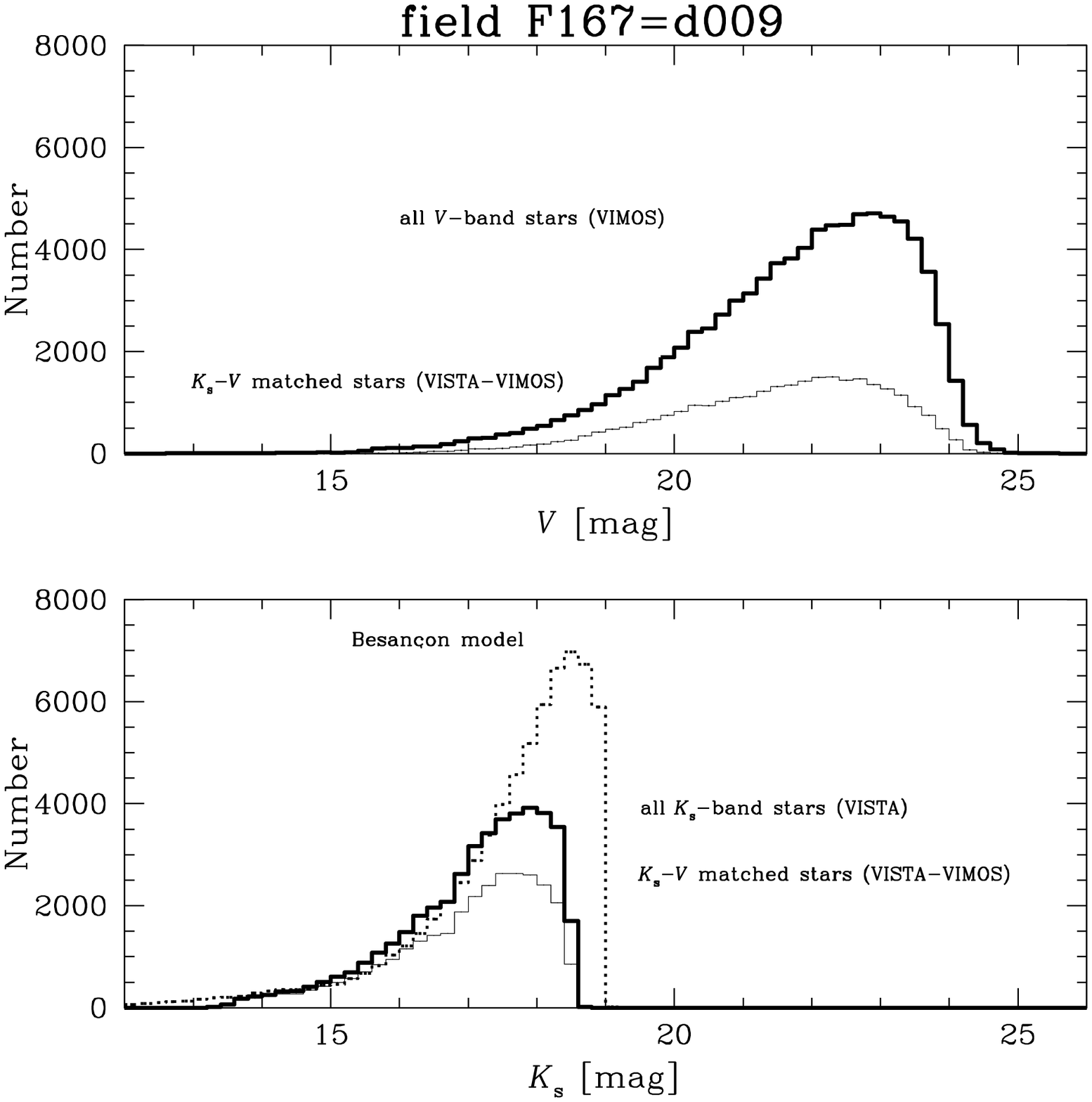}
\caption{Star count histograms in the $V$- (upper panel) and $K_s$-bands
(lower panel). The thin-line profiles show common VIMOS-VVV stars.
The Besan\c{c}on Galactic model is represented by the dotted line.
Note that around $K_s=17.5$~mag the model departs from the real data.}
\label{fig:histstarsF167}
\end{figure}

\begin{figure}
\includegraphics[angle=0,width=0.5\textwidth]{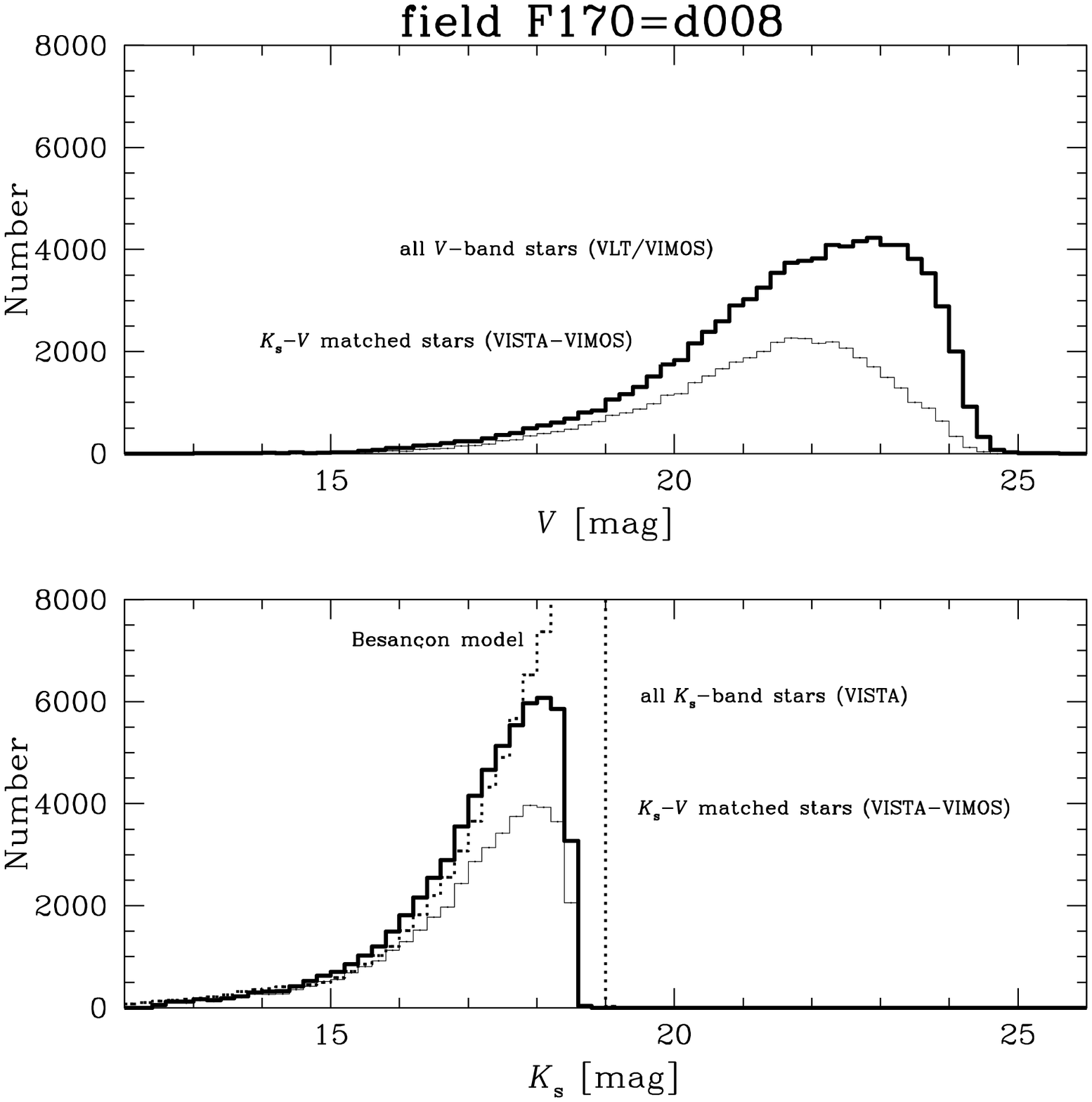}
\caption{Same as in Fig.~\ref{fig:histstarsF167} but for field F170.}
\label{fig:histstarsF170}
\end{figure}

Figs.~\ref{fig:histstarsF167} and~\ref{fig:histstarsF170} illustrate
the star count histograms for fields F167 and F170, respectively.
The histograms clearly show that the number of stars in the $K_s$-band
increases more rapidly than the one in the $V$-band. From the
location of maxima in the $V$-band (upper panels) one can find that
the optical data from VIMOS are about 1~mag deeper than the near-IR
data from VISTA. In the lower panel of Fig.~\ref{fig:histstarsF170}
we add a profile based on the Besan\c{c}on Galactic model, which we
describe in detail in Sec.~6. From the comparison of the models
with the observations we can infer that the analysed data are
complete down to $K_s\sim17.5$~mag.

\subsection{Colour-magnitude and colour-colour diagrams}

\begin{figure}
\includegraphics[angle=0,width=0.5\textwidth]{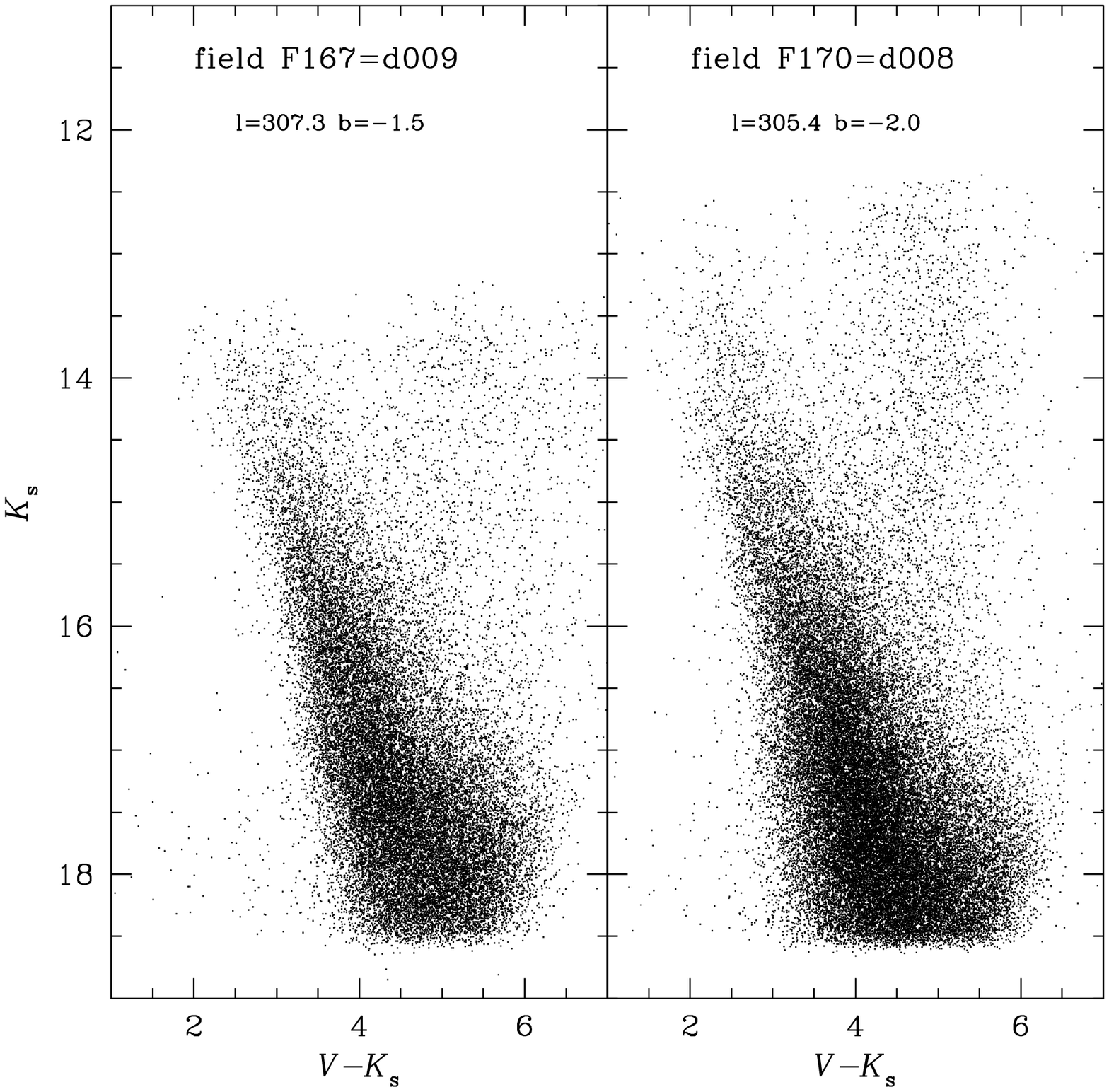}
\caption{$K_s$ vs. $V-K_s$ diagrams for common VIMOS-VVV
stars in the VIMOS fields F167 (left panel) and F170 (right panel).}
\label{fig:KVKcmd}
\end{figure}

In Fig.~\ref{fig:KVKcmd} we show $K_s$ vs. $V-K_s$ CMDs for the
two observed VIMOS fields. The main features in both diagrams are
the main sequence and the area of red giants. This is particularly
obvious in field F170, where the saturation level is higher.
Significant distance and reddening effects can be easily noted.
The red giants spreads over at least 2~mag in the $K_s$-band
($12.5<K_s<14.5$~mag) and 1~mag in colour ($4.5<V-K_s<5.5$~mag).
According to the newest view of our Galaxy \citep{chu09},
most observed stars in these two fields are very likely objects
located in the Scutum-Centaurus Arm, which is tangent to the
line of sight in this direction. Distances of these stars
are very likely in the range from 4 to 10~kpc. Assuming a maximum
reddening of $E(B-V)\sim2.1$~mag in the investigated disc fields
\citep{sch98}, and a 'universal' extinction law $A_V=3.1 E(B-V)$,
we obtain a maximum absorption of $A_V\sim6.5$~mag.

Fig.~\ref{fig:KVKcmd} also shows a lack of faint clump giants (with $K<15$~mag),
which can be interpreted as the line of sight leaving the stellar
edge of the Milky Way plane \citep{min11}.

\begin{figure*}
\includegraphics[angle=0,width=1.0\textwidth]{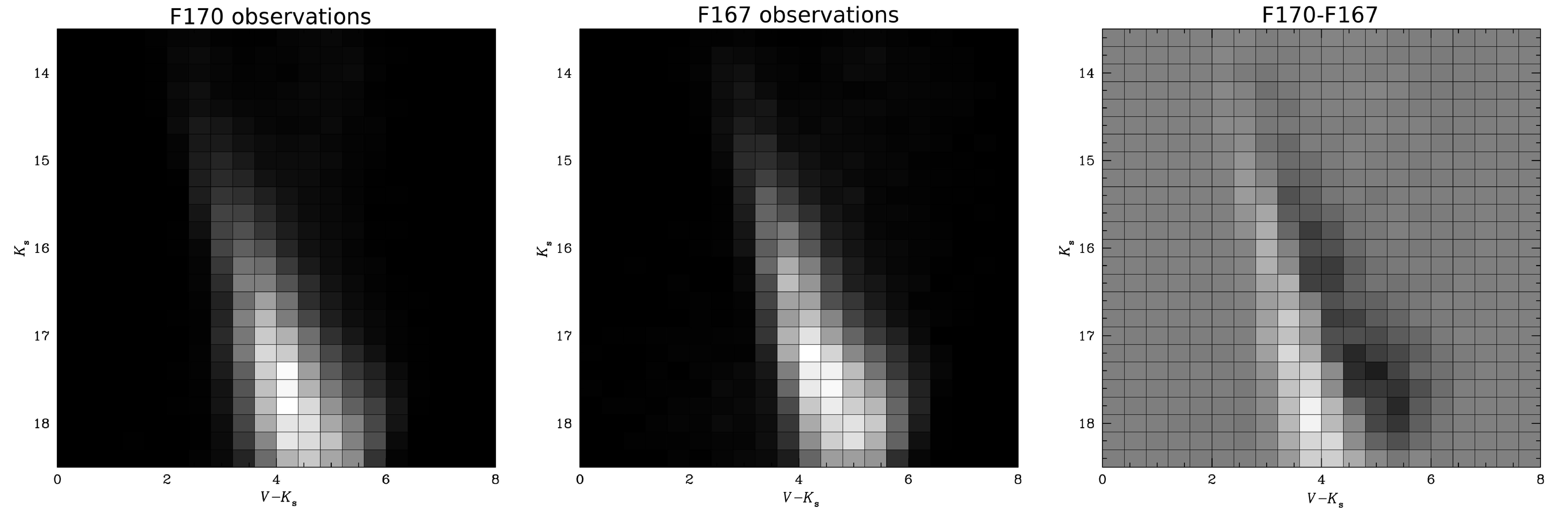}
\caption{Hess $K_s$ vs. $V-K_s$ diagram for the field F170 (left panel)
and F167 (middle panel). The subtraction of these two diagrams is shown
in the right panel.}
\label{fig:cmdF170mF167small}
\end{figure*}

For the two VIMOS fields we have also constructed Hess diagrams using
a bin size of 0.2~mag in $K_s$ and 0.4~mag in $V-K_s$, and shown
in Fig.~\ref{fig:cmdF170mF167small} together with the subtraction of the
F167 diagram from that of F170. The results confirm that the reddening
in field F167 is higher than in F170, as it is to be expected
because of its slightly closer position to the Galactic plane
($b_{\rm F170}\approx-2.0\degr$ vs. $b_{\rm F167}\approx-1.5\degr$).

\begin{figure}
\includegraphics[angle=0,width=0.5\textwidth]{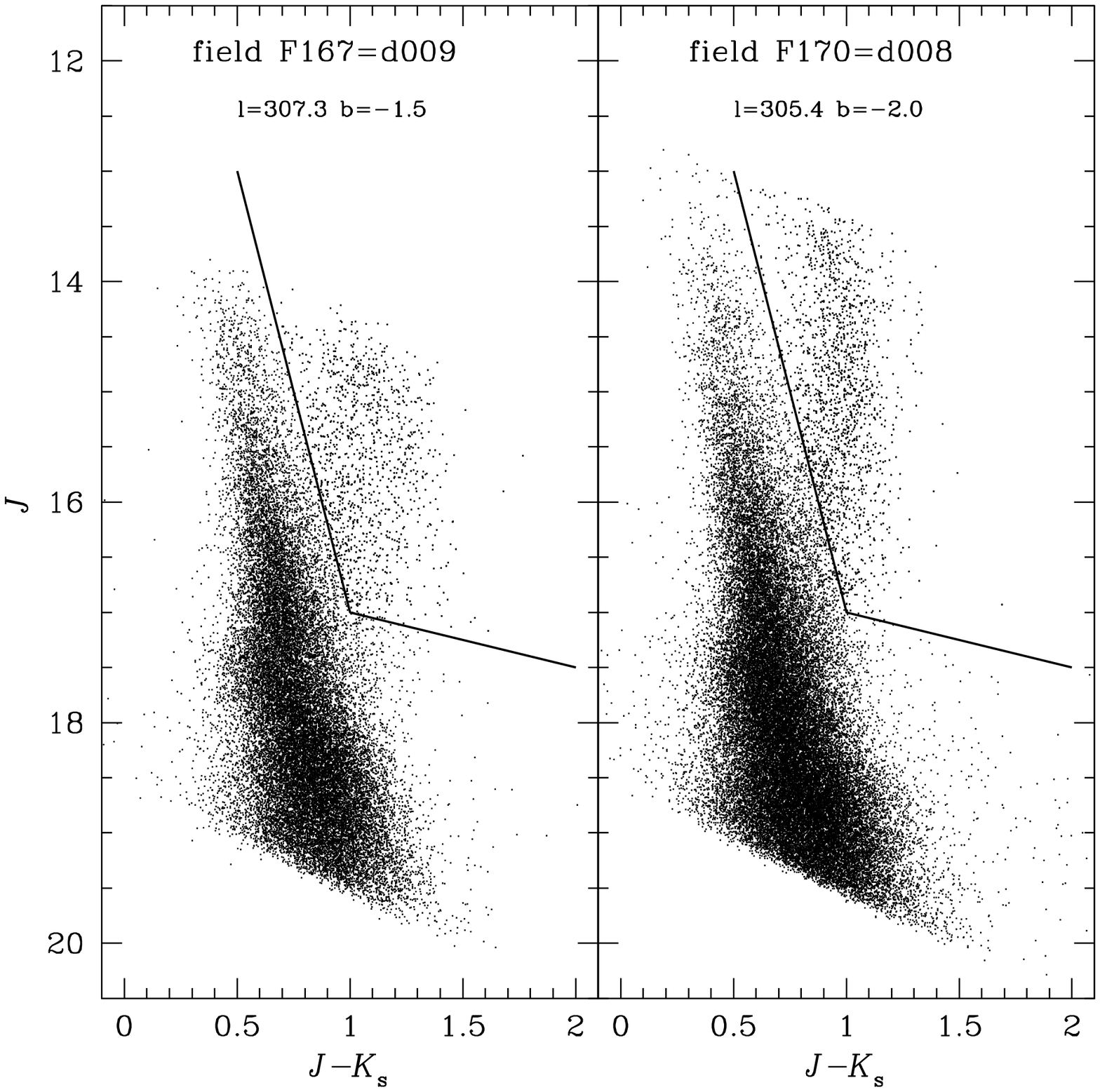}
\caption{$J$ vs. $J-K_s$ colour-magnitude diagrams for common VIMOS-VVV
stars in the VIMOS fields F167 (left panel) and F170 (right panel).
The solid lines roughly separate main-sequence from red giant stars.}
\label{fig:JJKcmd}
\end{figure}

\begin{figure}
\includegraphics[angle=0,width=0.5\textwidth]{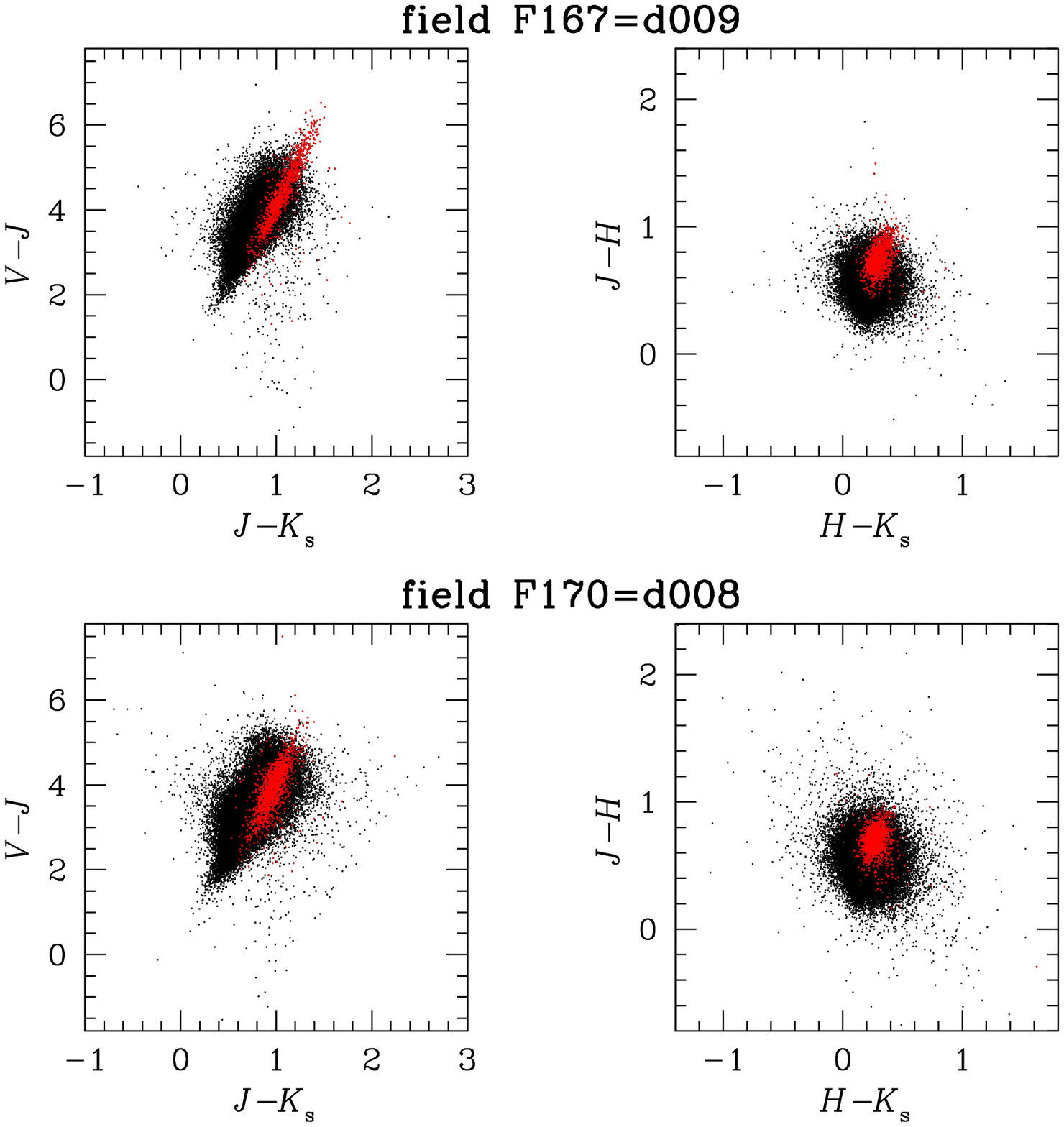}
\caption{Two ($V-J$ vs. $J-K_s$ and $J-H$ vs. $H-K_s$) colour-colour diagrams
for common VIMOS-VVV stars in the VIMOS fields F167 and F170.
Red giants selected in the CMDs in the previous figure are shown in red.}
\label{fig:colours}
\end{figure}

Fig.~\ref{fig:JJKcmd} shows $J$ vs. $J-K_s$ CMDs for the two VIMOS fields.
Here we arbitrarily separated red giants from main sequence stars with
a solid line. The two luminosity classes are also marked in the colour-colour
diagrams presented in Fig.~\ref{fig:colours}.

\begin{figure}
\includegraphics[angle=0,width=0.5\textwidth]{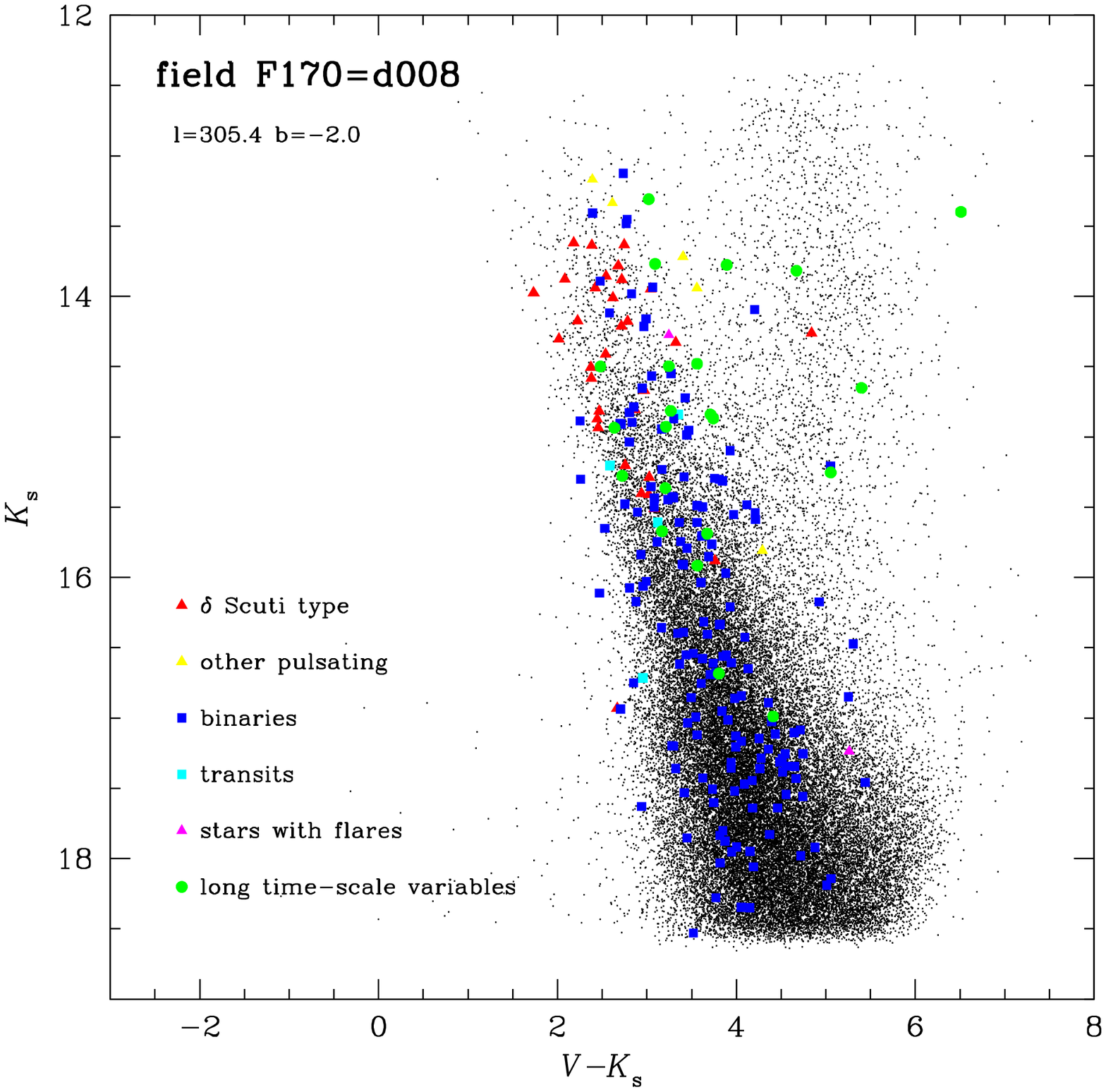}
\caption{$K_s$ vs. $V-K_s$ diagram with marked positions
of 223 variables found in the VIMOS field F170.}
\label{fig:d008KVK}
\end{figure}

Of the 333 detected variables in field F170 we have obtained $K_s$-band
magnitudes for 223 with VISTA. We marked the positions of these objects
in the $K_s$ vs. $V-K_s$ diagram presented in Fig.~\ref{fig:d008KVK}.
We stress that their positions in the CMD are still uncertain,
because $K_s$-band light curves and consequently average magnitudes
are not available yet. At this point we are able to infer
the following: (1) probably all observed eclipsing and ellipsoidal
variables are main-sequence binaries; (2) four transiting objects
are main-sequence stars; (3) most of the detected $\delta$-Scuti pulsators
seem to be located in the closer part of the Scutum-Centaurus Arm.
The last point results from the fact that $\delta$-Scuti variables
have low amplitudes of $0.01-0.12$~mag. This type of stars placed in the arm
should have $13<V<24$~mag (assuming $A_V<6.5$~mag), whereas we have
detected $\delta$-Scuti pulsators with $15.7<V<19.6$~mag.

\section{Comparison with the Galactic model of Besan\c{c}on}

We have compared our data with the stellar population synthesis model
of our Galaxy developed by the Besan\c{c}on group \citep{rob03}.
For the purpose of this work we have simulated a model assuming a total
integration time of 10~Gyr, constant star-formation rate, and no
kinematic effects. This was done for stars of absolute
brightness $-7<M_V<+20$~mag, spectral types between O0 and D5,
luminosity classes I to VI, and distances up to 50~kpc from the
Galactic center. The selected sky region in our study includes
the thin disc, thick disc, and the spheroidal halo.
In the Besan\c{c}on model the disc truncates at 14.0~kpc
from the Galactic centre located at a distance of 8.5~kpc from the Sun.
The profile densities of stars and interstellar matter are described
by the Einasto law. We have tested models with different values of
diffuse extinction $a_V$ and additional discrete clouds of a given
absorption $A_V$. The model output is a list of synthetic stars with information
on distance, absolute magnitude, luminosity class, age, mass, metallicity,
and observed magnitudes and colours.

\begin{figure*}
\includegraphics[angle=0,width=1.0\textwidth]{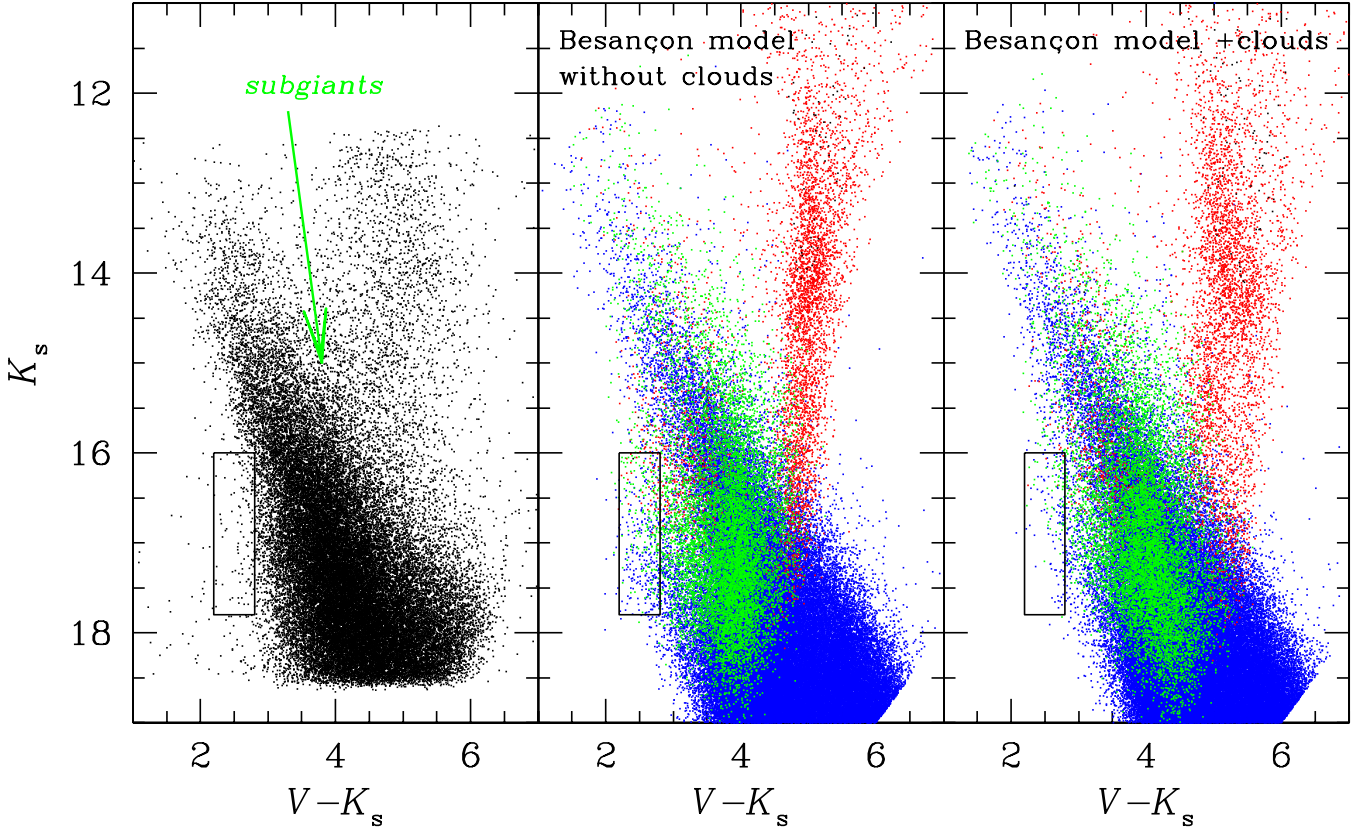}
\caption{Comparison of the observed (left panel) and two modelled
$K_s$ vs. $V-K_s$ diagrams for the field F170. Stars of different
luminosity classes are shown in colours: supergiants (class I)
and bright giants (class II) in black, giants (class III) in red,
subgiants (class IV) in green, main-sequence stars (class V) in blue.
Both models are for the standard extinction $a_V=0.70$~mag/kpc,
but the one in the right panel contains additional interstellar clouds
at distances 11-13~kpc from the Sun. The wider spread in the $V-K_s$
colour for giants and proper number of stars in the small box
indicate that the model with the clouds provides the better fits.}
\label{fig:obsvsmodelF170}
\end{figure*}

In Fig.~\ref{fig:obsvsmodelF170} we compare two modelled $K_s$ vs. $V-K_s$
diagrams with the observed one for field F170. The colours in the synthetic
CMDs represent different luminosity classes. The first of the modelled diagrams
(in the middle panel) is for the standard value of extinction $a_V=0.70$~mag/kpc
and without clouds. In general the observations and this model agree well,
indicating that the model is not missing any important components.
For example, based on the model, we can recognize subgiant stars in the observed
CMD around $K_s\sim15.5$~mag and $V-K_s\sim4.0$~mag.

\begin{figure*}
\includegraphics[angle=0,width=1.0\textwidth]{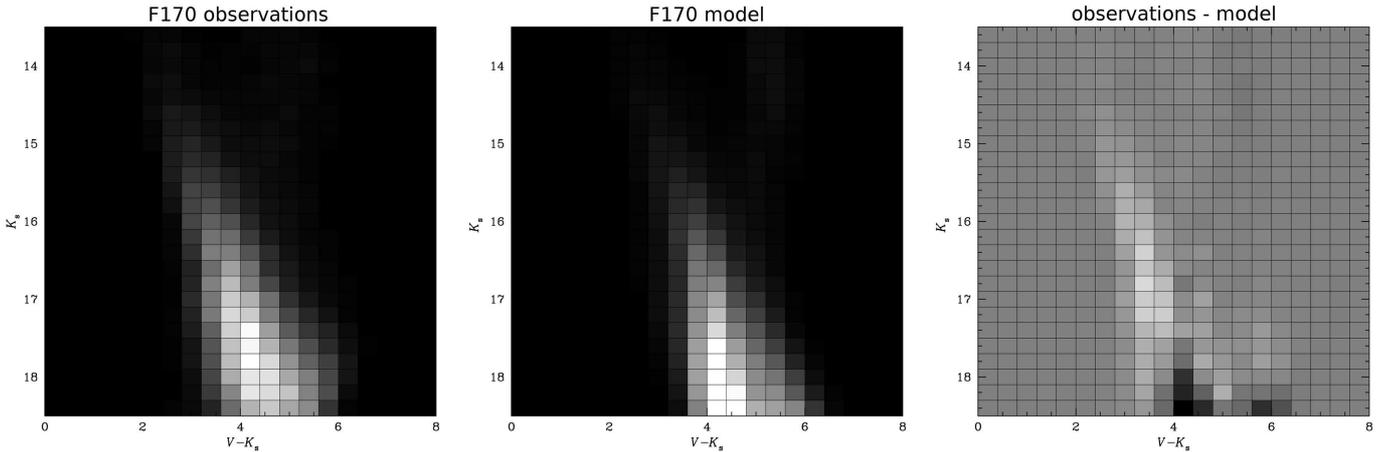}
\caption{Observed (left) and modelled (middle) Hess $K_s$ vs. $V-K_s$ CMDs.
The subtraction of these two diagrams is shown in the right panel.
For objects fainter than $K_s\sim17.5$~mag the data are incomplete.}
\label{fig:cmdF170OmTsmall}
\end{figure*}

There are also differences between the first model and observations. The most
noticeable difference is the spread in the $V-K_s$ colour for giants,
which is much wider in the observational diagram. This effect partially
stems from significant differential reddening in the observed field, but
can also be improved by adding interstellar clouds.
One can also see too many simulated stars within the area between
$16.0<K_s<17.8$~mag and $2.2<V-K_s<2.8$~mag (see boxed section in all
panels of Fig.~\ref{fig:obsvsmodelF170}), which are actually not observed.
In the first modelled CMD this area is populated by giants, subgiants,
and luminous main-sequence stars ($-0.5<M_V<2.0$~mag) of relatively
high metallicities ($-0.5<[Fe/H]<0.3$) located at distances $10<D<18$~kpc.
We tested models with different values of extinction and
interstellar clouds beyond 10~kpc. The best result we achived for the
standard extinction $a_V=0.70$~mag/kpc and three absorbing clouds of
$a_V=0.17$~mag each located at distances 11, 12, and 13~kpc from
the Sun, respectively. These clouds would exist in the Carina-Sagittarius
Arm, which is a minor Galactic arm located between the Scutum-Centaurus Arm
and the distant extended spiral arm of the Milky Way \citep{mcc04}.
This model is presented in the right panel of Fig.~\ref{fig:obsvsmodelF170}.
In Fig.~\ref{fig:cmdF170OmTsmall} we show the result of subtraction
of the observed and best-modelled CMDs.

\begin{figure*}
\includegraphics[angle=0,width=1.0\textwidth]{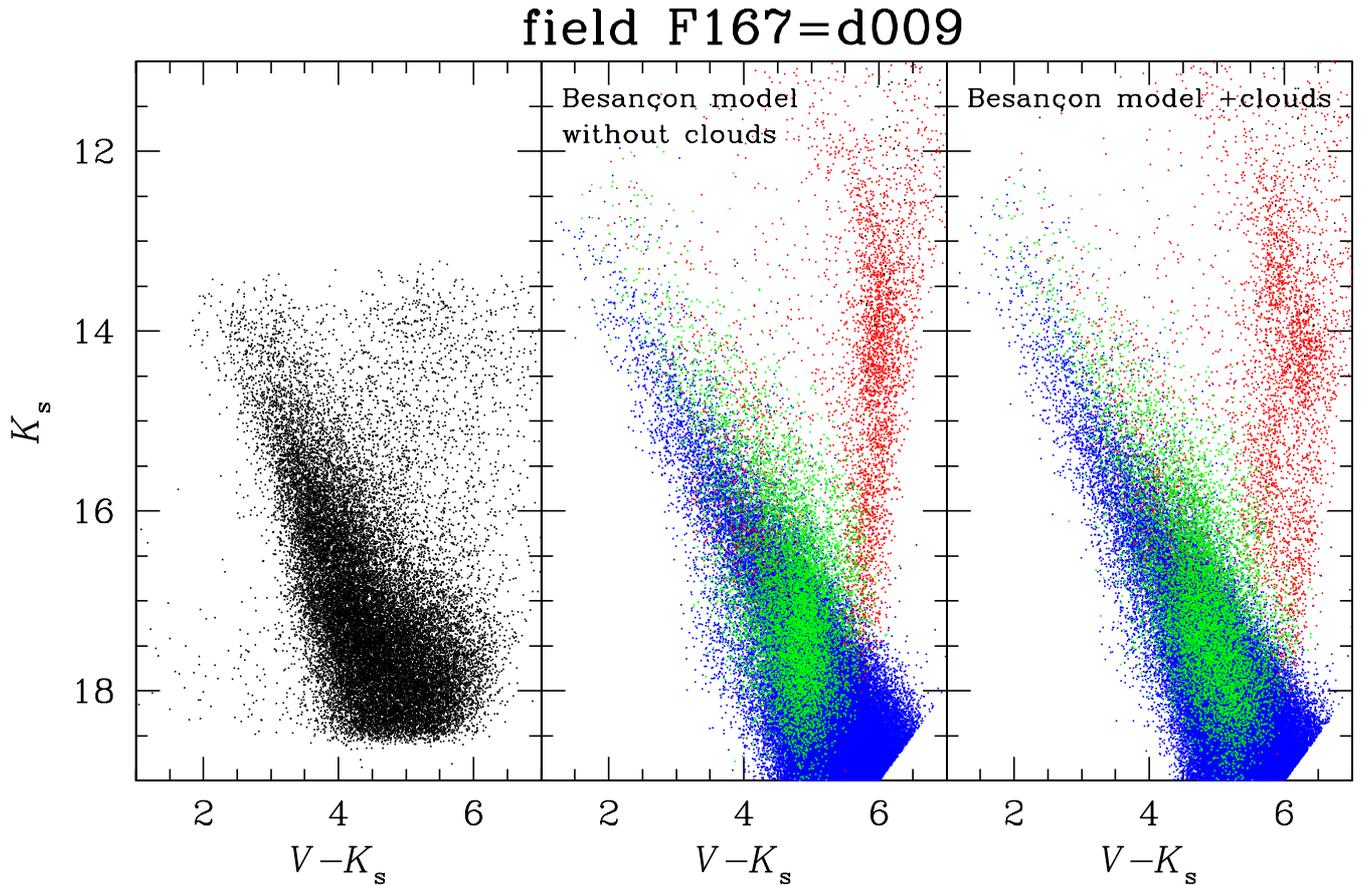}
\caption{Comparison of the observed (left panel) and modelled
$K_s$ vs. $V-K_s$ diagrams for field F167.
The colours are the same as in Fig.~\ref{fig:obsvsmodelF170}.
The model in the middle panel is for the standard extinction
$a_V=0.70$~mag/kpc and does not contain absorbing clouds. The model
in the right panel is for a lower extinction of 0.68~mag/kpc and contains
clouds located at the same distances as in the best model for F170,
i.e. at 11-13~kpc from the Sun.}
\label{fig:obsvsmodelF167}
\end{figure*}

Results of simulations for the field F167 are presented
in Fig.~\ref{fig:obsvsmodelF167}. In this case we left the clouds,
but we had to decrease the extinction to $a_V=0.68$~mag/kpc.

\section{Conclusions}

We have used deep optical data from VLT/VIMOS and near-IR data from VISTA
to search for variable stars and to investigate stellar populations near
the Galactic plane in the constellation Centaurus. In the VIMOS field
F170 in a series of 583 $V$-band images spread over two nights we have detected
333 variables among 84~734 stars with brightness in the range $12.7<V<26$~mag.
Only four of these objects were previously known.
For 263 variables we were able to assess the periods.
Fifty-three of the newly discovered variables are pulsating stars,
mostly of $\delta$~Scuti type. Thirty-eight objects show brightness
variations on long-time scales, and require a longer timeline of observations
to derive their period and mean luminosity.

\begin{table*}[htb]
\caption{Percentage of detected binaries in our search in comparison
with other ground-based surveys and recent results from the {\it Kepler}
space mission (the last row).}
\smallskip
{\small
\begin{tabular}{lcccrc}
\hline
Authors & Field & Period of obs. & Method of searches & Total number of stars & Percentage of ecl/ell \\
\hline
Paczy\'nski et al. (2006)  & $\delta<+28\degr$    & 5-8 years  &       Fourier & 17,000,000 & 0.065 \% \\
Weldrake \& Bayliss (2008) & 0.751~deg$^2$, Lup   &  53 nights &           AoV &    110,372 & 0.172 \% \\
Pietrukowicz et al. (2009) & 0.052~deg$^2$, Car   & ~~4 nights & by eye \& AoV &     50,897 & 0.291 \% \\
Miller  et al. (2010)      & 0.25~deg$^2$, Nor    &  58 nights &           LSA &    335,592 & 0.301 \% \\
This work                  & 0.052~deg$^2$, Cen   & ~~2 nights &        by eye &     84,734 & 0.284 \% \\
\hline
Pr\v sa et al. (2011)      & 105~deg$^2$, Cyg-Lyr & ~~44 days  &           BLS &    156,097 & 1.204 \% \\
\hline
\label{tab:binaries}
\end{tabular}}
\end{table*}

Among all observed variables 241 are classified as eclipsing/ellipsoidal
binaries, which is 0.284\% of the whole sample of searched stars.
In Table~\ref{tab:binaries} we compare this result with the rate of binaries
detected in Galactic fields in other recent ground-based variability studies
as well as in the {\it Kepler} space mission. The simple search method
we applied to our short-period data, the eye inspection of all light
curves, turned out to be very effective. It is interesting that
we have obtained almost the same rate of eclipsing/ellipsoidal
systems as in our previous search in a VIMOS field in Carina
\citep{pie09}. The binary detection rate would be very probably higher
if the observations lasted considerably longer than only two days.
Preliminary results from the {\it Kepler} space mission show
that after the first 44 days of operation the average
occurrence rate of binary systems is $\sim1.2$\% \citep{prsa11}.
However, the {\it Kepler} photometry is much more accurate, which allows
the detection of a larger part of lower amplitude variables than the
other ground-based surveys. About 51\% of 1879 eclipsing/ellipsoidal
binary stars detected by {\it Kepler} have orbital periods shorter than
1.44~days, whereas in the VIMOS sample 203 out of 241 binary systems have
orbital periods $<1.44$~d. Assuming the same binary occurrence rate in all
directions of the Milky Way we estimate that one can find up to $\sim0.5$\%
of binaries in ground-based data using an image subtraction technique.

In the second part of this work we have combined the $V$-band data
from VIMOS with $JHK_s$-band photometry from VISTA.
This has been done for the VIMOS fields F167 and F170. We
presented optical-IR colour-magnitude and colour-colour diagrams.
By subtracting CMDs for field F167 from field F170 we confirmed
that the reddening in field F167, which is located slightly
closer to the Galactic plane, is indeed higher than in F170.
From the location of the detected variables in the CMD we conclude
that probably all objects belong to main-sequence stars in the
Scutum-Centaurus Arm.

From the comparison of the observed $K_s$ vs. $V-K_s$ diagrams with
different synthetic diagrams based on the Galactic model of Besan\c{c}on
we conclude the presence of interstellar clouds at distances 11-13~kpc
from the Sun. The clouds may belong to the minor Carina-Sagittarius
Arm located between the Scutum-Centaurus Arm and the distant
extended spiral arm of the Milky Way.

The results presented in this paper, including the large number of newly
discovered variables in a relatively small field, high-quality photometry,
and the construction of deep optical-IR colour-magnitude diagrams
demonstrate that ground-based wide-field multiple-band variability
surveys are powerful tools for drawing a detailed picture of our Galaxy.

\begin{acknowledgements}
PP is supported by funding to the OGLE project from the European
Research Council under the European Community$'$s Seventh Framework
Programme (FP7/2007-2013)/ERC grant agreement no. 246678, and
by the grant No. IP2010 031570 financed by the Polish Ministry of
Sciences and Higher Education under Iuventus Plus programme.
This work is also supported by FONDAP Center for Astrophysics 15010003,
BASAL Center for Astrophysics and Associated Technologies PFB-06,
MILENIO Milky Way Millennium Nucleus P-07-021-F, FONDECYT 1090213
from CONICYT, and the European Southern Observatory.
\end{acknowledgements}

\end{document}